\documentclass{article}
\usepackage[ruled,vlined]{algorithm2e}
\usepackage{spconf,amsmath,graphicx,hyperref}
\usepackage{cite}
\usepackage{amsmath,amssymb,amsfonts}
\usepackage{algorithmic}
\usepackage{textcomp}
\usepackage{xcolor}
\usepackage[table]{xcolor}
\usepackage{multirow}
\usepackage{booktabs}
\usepackage{cleveref}
\usepackage{bigdelim} 
\hypersetup{
  colorlinks=true,    
  linkcolor=black,     
  citecolor=black,    
  urlcolor=black 
}

\title{FakeSound2: A Benchmark for Explainable and Generalizable Deepfake Sound Detection}

\name{Zeyu Xie$^{1}$, Yaoyun
 Zhang$^{2}$, Xuenan Xu$^{2}$, Yongkang Yin$^{1}$, Chenxing Li$^{3}$, Mengyue Wu$^{2}$\textsuperscript{*}, Yuexian Zou$^{1}$\textsuperscript{*}\thanks{* Corresponding authors}}

\address{$^{1}$ Guangdong Provincial Key Laboratory of Ultra High Definition Immersive Media Technology, \\
Peking University, Shenzhen \\ 
         $^{2}$ Shanghai Jiao Tong University, Shanghai
         $^{3}$ Tencent AI Lab, Beijing
         }
\begin{document}
\ninept
\maketitle
\begin{abstract}
The rapid development of generative audio raises ethical and security concerns stemming from forged data, making deepfake sound detection an important safeguard against the malicious use of such technologies. 
Although prior studies have explored this task, existing methods largely focus on binary classification and fall short in explaining \textit{\textbf{how}} manipulations occur, tracing \textbf{\textit{where}} the sources originated, or \textit{\textbf{generalizing}} to unseen sources—thereby limiting the explainability and reliability of detection.
To address these limitations, we present \href{https://zeyuxie29.github.io/FakeSound2/}{\textcolor{cyan}{\textit{FakeSound2}}}\footnote{The resources are available at: \href{https://zeyuxie29.github.io/FakeSound2/}{\textcolor{cyan}{\textit{https://zeyuxie29.github.io/FakeSound2/}}}}, a benchmark designed to advance deepfake sound detection beyond binary accuracy. 
FakeSound2 evaluates models across three dimensions: localization, traceability, and generalization, covering 6 manipulation types and 12 diverse sources.
Experimental results show that although current systems achieve high classification accuracy, they struggle to recognize forged pattern distributions and provide reliable explanations. 
By highlighting these gaps, FakeSound2 establishes a comprehensive benchmark that reveals key challenges and aims to foster robust, explainable, and generalizable approaches for trustworthy audio authentication.

\end{abstract}
\begin{keywords}
AIGC, deepfake sound detection, explainability, traceability,  generalization
\end{keywords}

\section{Introduction}
\label{sec:intro}

The rapid advancement of generative artificial intelligence has revolutionized the creation of synthetic media, enabling the generation of highly realistic synthetic content.
However, the malicious use of generative media can raise serious ethical and legal concerns, making the development of effective deepfake detection imperative.

\begin{figure}[t]
\centerline{\includegraphics[width=1.0\linewidth]{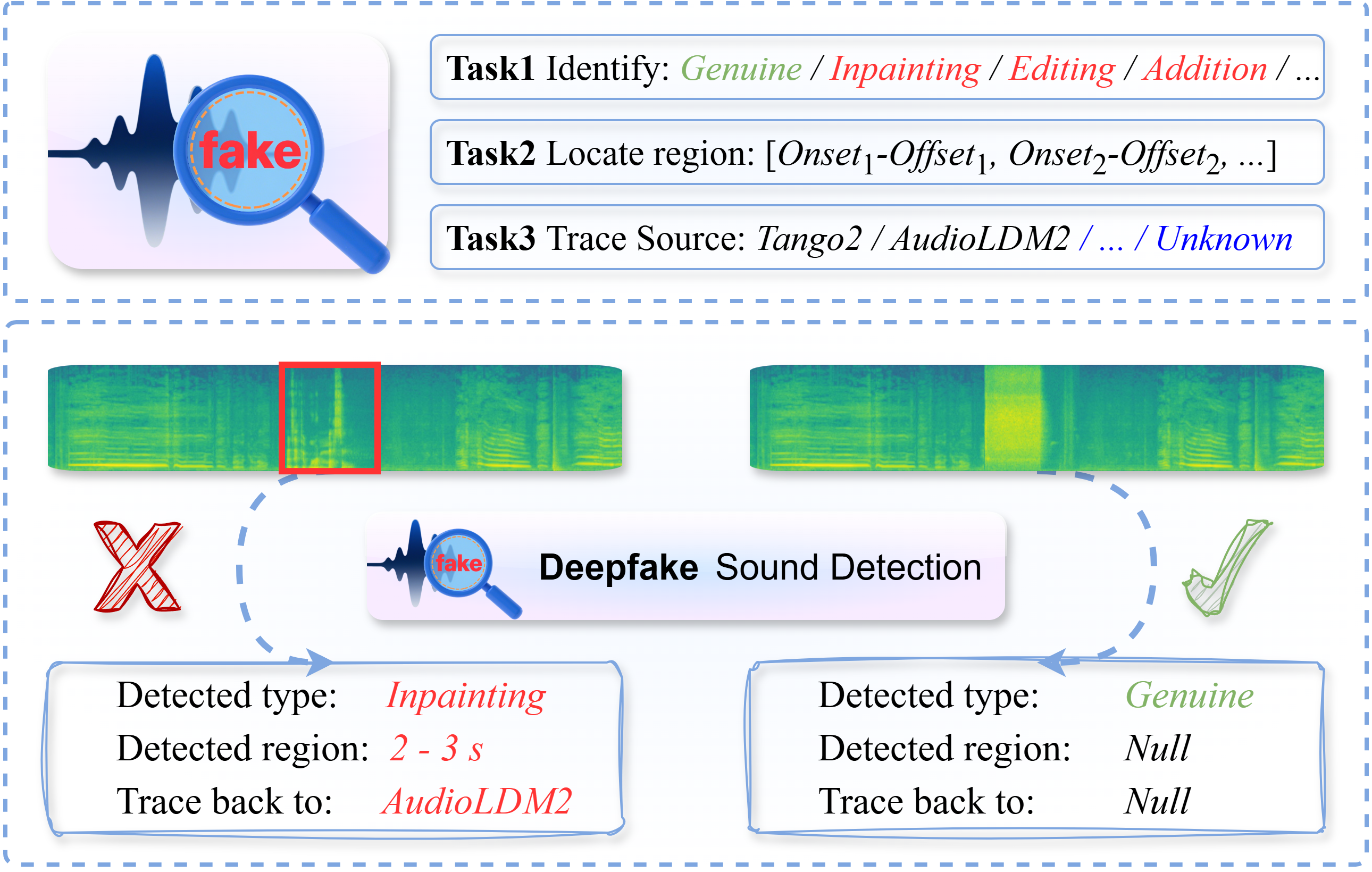}}
\caption{
The explainable deepfake detection need identify the manipulation method (\textit{\textbf{how}}), localize the temporal positions of forgery (\textit{\textbf{when}}), and trace the accountable sources (\textit{\textbf{where}}).
}
\label{fig:sample}
\end{figure}

While deepfake detection in the visual~\cite{dolhansky2020deepfake} and speech~\cite{yi2023add, yamagishi2021asvspoof} domains has received considerable attention, the field of Deepfake Sound Detection (DSD) remains markedly understudied. 
The recent advanced audio generation models~\cite{hai2024ezaudio, hung2024tangoflux, shan2025hunyuanvideo} are capable of producing highly realistic sounds indistinguishable from genuine recordings, which poses significant risks of malicious exploitation such as fabrication of evidence and news fabrication. 
This underscores the critical importance of DSD technology in mitigating social and ethical risks.
Although several studies have undertaken preliminary explorations in DSD \cite{yin2025envsdd, fu2025rpra}, they focus on simple binary classification at the clip level. 
Such approaches fail to provide temporal localization of forgeries, distinguish between manipulation types, or trace the source of synthetic content, thereby significantly limiting the practical utility and forensic value of the detection system. 
In real-world scenarios, however, explainability and traceability are critical not only for building trustworthy detection systems but also for understanding the nature of an attack, attributing forgeries to specific generators or sources, and ultimately supporting human decision-making in security-sensitive applications.

To this end, we introduce a comprehensive benchmark FakeSound2 to evaluate the localization, interpretability, and generalization capability of DSD models, as shown in Figure~\ref{fig:sample}.
Using FakeSound2, we systematically identify limitations in existing approaches. 
Evaluation results demonstrate that, although current models achieve excellent performance on binary real / fake classification, they exhibit notable deficiencies in explaining manipulation types and sources, as well as in generalizing to unseen data.
This indicates that current models have not truly learned the intrinsic nature of forgeries but merely rely on empirical patterns in training data. 
It advocates for a new research direction in deepfake sound detection, one that prioritizes explainability and generalization as fundamental pillars of reliable and trustworthy forensic analysis.

\begin{table*}[htbp]
    \centering
    \caption{Metadata of FakeSound2. 
    ``Test-F" denotes filtered test subsets. 
    The X2Audio model is an in-house trained text-to-audio system. }
    \label{tab:inter}
    
    \centering
    \begin{tabular}{c|l|l|c|l|c}
        \toprule
        \textbf{Type} & \multicolumn{1}{c|}{\textbf{Manipulation }} &  \multicolumn{1}{c|}{\textbf{Source}} & \textbf{Num of Train / Test / Test-F} &\multicolumn{1}{c|}{\textbf{Unseen source}} & \textbf{Num of Test / Test-F} \\
        \midrule
        \multirow{3}{*}{Clip -wise} & 0. Genuine & - & 49501 / 964 / 964 & - & -\\
        &  1. Generation & Affusion\cite{xue2024auffusion} & 49501 / 964 / 498 & X2Audio (In-house) & 964 / 426  \\
        &  2. Editing & AudioEditor\cite{jia2025audioeditor} & 9059 / 645 / 274 & Audit\cite{wang2024audit} & 645 / 32 \\
        \midrule

       \multirow{7}{*}{Frame -wise}& \multirow{3}{*}{3. Inpainting}\rdelim\{{3}{1em}    &Tango2\cite{majumder2024tango}   & 49501 / 964 / 531&   \multirow{2}{*}{\rdelim\{{2}{1em} MakeAnAudio\cite{huang2023make1}} & \multirow{2}{*}{964 / 452} \\
        
        &  & \multicolumn{1}{l|}{Affusion\cite{xue2024auffusion}} & 49501 / 964 / 582& \hspace{0.8em} \multirow{2}{*}{FakeSound\cite{xie2024fakesound}} & \multirow{2}{*}{632 / 632}  \\
        &  & \multicolumn{1}{l|}{AudioLDM2\cite{liu2023audioldm2}} & 49501 / 964 / 470 & &  \\
        & 4. Separation  & LASSNet\cite{liu2022separate} & 34410 / 645 / 102 & FlowSep\cite{yuan2025flowsep} & 645 / 115 \\
        & 5. Splicing  & Genuine& 31013 / 591 / 206   & - & -\\
        & 6. Addition& Genuine&  47942 / 674 /  269 & - & -\\
        \midrule
        Total & & &369929 / 7375 / 3896& & 3850 / 1657 \\
        \bottomrule
    \end{tabular}
    
\end{table*}

\begin{figure*}[t]
\centerline{\includegraphics[width=0.75\linewidth]{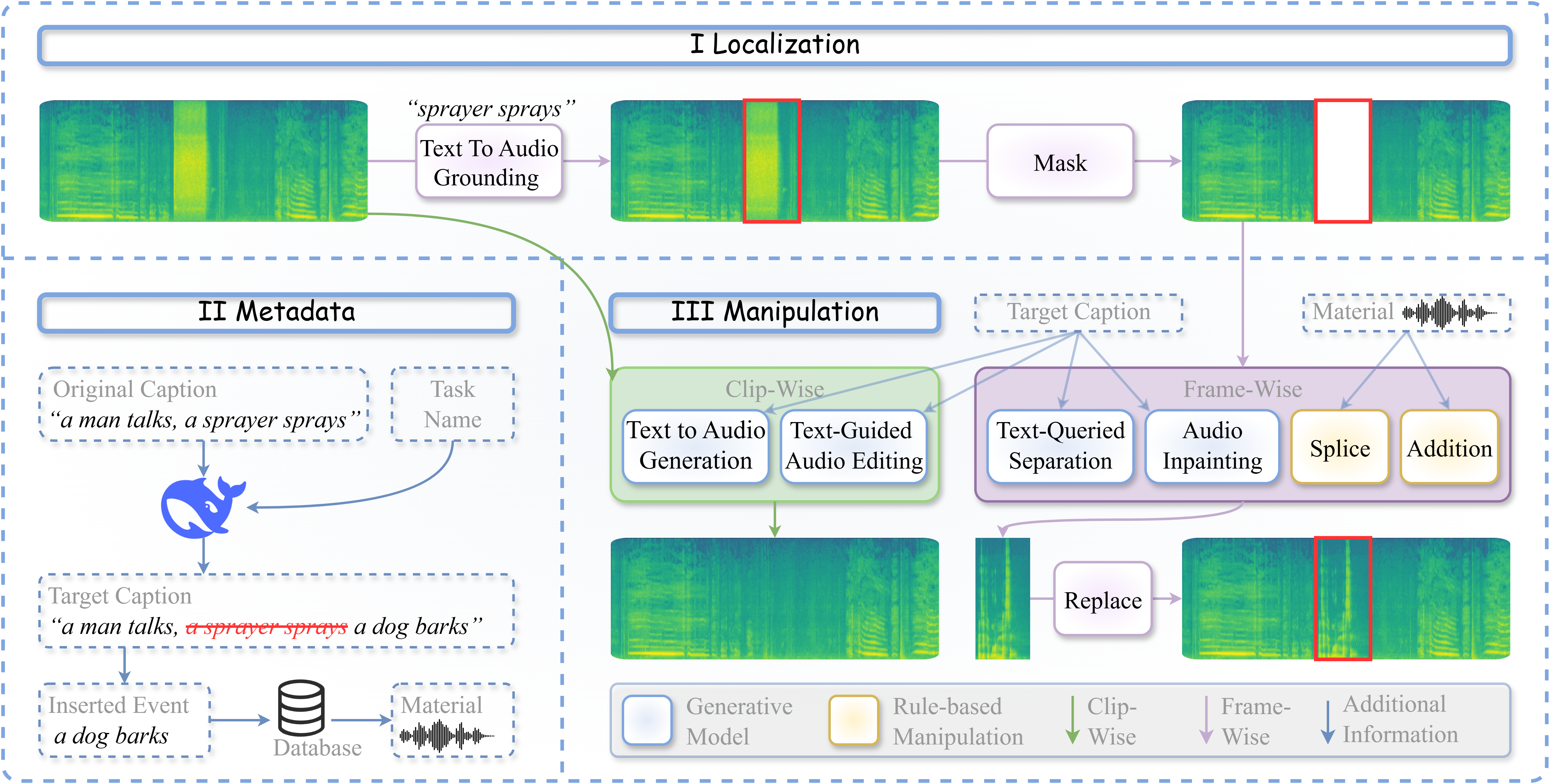}}
\caption{
Manipulation pipeline: 
    (I) A Text-to-Audio grounding model temporally localizes sound events and masks segments;
    (II) Based on the task type, the LLM generates the target captions and insert event descriptions;
    (III) Depending on the task requirements, relevant models or scripts are invoked to generate synthetic audio segments; frame-wise forgeries are seamlessly spliced according to the masked regions.
}
\label{fig:pipeline}
\end{figure*}


\begin{figure*}[tbp]
\centerline{\includegraphics[width=1\linewidth]{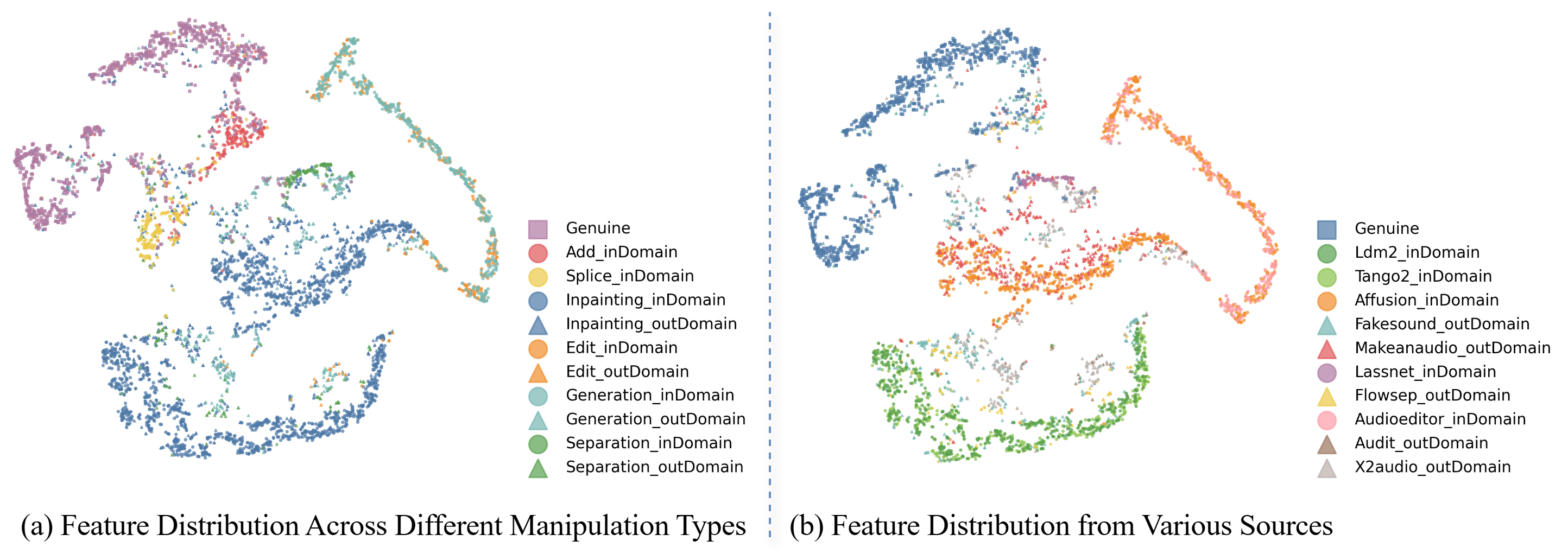}}
\caption{
The feature distribution of deepfake sound detection model.
Overall, the current model exhibits reliable separability between authentic and manipulated audio, enabling effective binary classification. 
However, certain forged audio categories demonstrate substantial overlap in the feature space, which diminishes explainability for source attribution.
}
\label{fig:feature_visualization}
\end{figure*}


\section{Explainable Deepfake Detection Definition}
\label{sec:method}
Establishing explainability in deepfake sound detection necessitates a systematic classification of potential audio manipulation techniques.
As outlined in Table~\ref{tab:inter}, we surveyed prevalent methods and temporally divided them into clip-wise manipulation, wherein the entire audio clip is synthetic, and frame-wise manipulation, where only certain selected frames are altered. 
Clip-wise approaches include (1) \textbf{\textit{Generation}}, which synthesizes an audio clip conditioned on an caption, and (2) \textbf{\textit{Editing}}, which modifies a given audio clip according to a target caption. 
Frame-wise manipulations comprise (3) \textbf{\textit{Inpainting}}, where masked frames are reconstructed based on a caption; 
(4) \textbf{\textit{Separation}}\footnote{Although separation is commonly associated with editing tasks, we treat it as a distinct category due to its reliance on task-specific models and its capacity to enhance perceptual authenticity through splicing with genuine audio segments.}, which removes events specified by a target caption and splices the remaining segments with genuine segments; 
(5) \textbf{\textit{Splicing}}, concatenating authentic clips guided by a caption; 
and (6) \textbf{\textit{Addition}}, where an external audio segment is inserted into an original clip according to a textual description.
Each manipulation method involves a distinct source of content, encompassing generative models (1)–(4) and authentic audio sources (5)–(6).

Based on the above observation, we posit that an effective deepfake sound detection model should possess the following key characteristics:
\underline{Localization}: not only discriminate between authentic and manipulated audio but also achieve precise temporal localization of forged segments in frame-wise manipulation scenarios.
\underline{Explainability}: analyze deepfake audio in terms of \textit{\textbf{how}} the audio was manipulated—by identifying the specific manipulation method and \textit{\textbf{where}} the manipulated content originated—through tracing the source of manipulated components.
\underline{Generalization}: maintain strong performance on unseen sources.

\section{FAKESOUND2 Construction}
\label{sec:dataset}

This section details the construction of the FakeSound2 dataset, a large-scale resource designed to address explainability and generalization in DSD. 
Constructed through an automated pipeline leveraging the AudioCaps~\cite{kim2019audiocaps} dataset, it encompasses \textbf{6} distinct manipulation types derived from \textbf{12} audio sources ($11$ synthetic and $1$ genuine).
As summarized in Table~\ref{tab:inter}, the dataset ultimately comprises a training set of 369,929 samples and a test set of 5,553 samples. 
Researchers can flexibly partition the test subset to evaluate generalization, provided that out-of-domain (OOD) sources remain completely unseen during training.
For instance, in our implementation, the test set is divided into 3,896 in-domain and 1,657 OOD samples to systematically benchmark cross-domain detection robustness.

As illustrated in Fig.~\ref{fig:pipeline}, the proposed pipeline comprises $4$ main stages:
(1) Event localization, where semantically rich key event segments are identified;
(2) Metadata processing, which employs large language models (LLMs) to generate target captions based on original captions and manipulation type;
(3) Audio manipulation, where generative models and rule-based scripts are applied to modify audio content;
(4) Quality filtering, in which low-quality segments are removed to ensure dataset reliability.
This structured approach facilitates the scalable creation of a high-quality and traceable dataset suitable for rigorous evaluation of deepfake sound detection systems. 

\subsection{Event localization}
\label{ssec:localization}
Starting from audio-caption pairs, we initially locate events in the audio, which are referred to as key segments. 
These segments, which carry rich semantic information, are essential for understanding the audio content, and any manipulation applied to these regions produces the most noticeable and impactful alterations\cite{xie2024fakesound}. 
Therefore, localizing the key segments is regarded as a critical starting point for deepfake detection.

The temporal boundaries (onset and offset) of sound event are localized by leveraging a Text-to-Audio grounding model~\cite{xu2024towards}. 
While for inpainting task, segments are randomly masked to enhance the model's generalization ability.


\subsection{Metadata Process}
\label{sec:metadata}
Leveraging the exceptional text-processing capabilities of LLM, we employ DeepSeek~\cite{liu2024deepseek} to simulate the human chain-of-thought process in audio manipulation, as illustrated in part II of Fig~\ref{fig:pipeline}. 
Specifically, the framework first parses the caption annotations into distinct event clauses; then, based on the task objective, it selectively removes existing events and inserts new ones.

For instance, the caption \textit{``a man talks and a sprayer sprays"} is partitioned into two separate clauses: \textit{``a man talks"} and \textit{``a sprayer sprays"}.
For both generation and inpainting tasks, the target caption remains consistent with the original, unaltered.
For the other tasks, it might remove \textit{``a sprayer sprays"} and insert \textit{``a dog barks"}.
The inserted events are selected from a library constructed from all parsed clauses and their corresponding localized audio segments.
%

\subsection{Audio Manipulation}
\label{ssec:manipulation}
To construct a dataset that addresses both explainability and traceability, we leverage a diverse range of generative models and rule-based methods, resulting in a comprehensive and richly annotated deepfake sound dataset.
Deepfake audio is synthesized by leveraging these models
and scripts, conditioned on the target caption.
All models are run using their default configurations.
In the frame-wise scenario, forged segments are spliced with authentic ones, resulting in a partially forged audio clip.


\subsection{Quality filtering}
\label{ssec:filter}
To ensure data quality, we conduct further filtering on the audio samples in the test set, resulting in a refined test-filtering (Test-F) subset. 
The Contrastive Language-Audio Pretraining (CLAP)~\cite{laionclap2023} model is utilized as the evaluation metric, with the similarity between the original audio and its corresponding caption serving as the benchmark. 
Any manipulated audio is discarded if its similarity to the target caption falls below the benchmark.

\begin{table*}[htbp]
    \centering
    \caption{Experimental result. 
    The white and \colorbox{gray!8}{light gray} backgrounds indicate clip-wise and frame-wise tasks, respectively.
    }
    \label{tab:result}
    
    \centering
    \begin{tabular}{c|cc|cc|c| cc}
        \toprule

        \multirow{2}{*}{\textbf{Task}} & \multicolumn{4}{c|}{\textbf{In domain}} & \multicolumn{3}{c}{\textbf{Out of domain}} \\
        \cline{2-8}
        
        & Acc$_{manipulation}$	& Acc$_{source}$	& Acc$_{identify}$ & F1$_{segment}$ &Acc$_{manipulation}$		& Acc$_{identify}$ & F1$_{segment}$ \\
        
      \midrule

      \multirow{1}{*}{Genuine} &83.09 &82.26 &86.41  &- & -& -&- \\ 
         
       \multirow{1}{*}{ Generation} &99.40	&99.60 &	100.00&100.00  & 4.23 &	86.62 &	81.60 \\ 
      
       \multirow{1}{*}{Editing} &87.96 &87.96 &100.00 &100.00  &0.00 &100.00 &91.42 \\ 

      \rowcolor{gray!8}\multirow{1}{*}{Inpainting} &99.75	 &93.68	 &99.94	 &98.91 &46.49 &71.86 &64.95 \\  

     \rowcolor{gray!8} \multirow{1}{*}{Separation} &93.14 &93.14	 &96.08	&94.14 &12.17	&96.52	&88.57 \\ 
 
      \rowcolor{gray!8}\multirow{1}{*}{Splicing} &90.78	 &94.66	 &89.32	 &87.77 & -& -& - \\ 

      \rowcolor{gray!8}\multirow{1}{*}{Addition} &85.13	 &88.85	 &87.73	&85.48  & -& -& - \\ 

      \rowcolor{gray!20} \multirow{1}{*}{Total} &93.10	 &90.91 &	95.10	&97.46	&32.35	&77.91	&74.66  \\

         
      

      





        
        \bottomrule
    \end{tabular}
    
\end{table*}

\section{Experiment}
\label{sec:model}
We employ FakeSound2 to evaluate the localization, explainability, and generalization of current models, thereby revealing their limitations and analyzing the fundamental challenges inherent in DSD tasks, while also establishing a benchmark for future research.


\subsection{Baseline}
The baseline model is build upon prior work~\cite{xie2024fakesound} and consists of an audio encoder and a backbone network, similar to Cai et al.~\cite{cai2023dku}. 
The audio encoder employs a high-performance self-supervised pre-trained model, EAT~\cite{chen2024eat}, to extract fine grain representations. 
The backbone architecture and parameters are identical to those in Cai et al.~\cite{cai2023dku}, comprising a 12-layer ResNet flanked by Convolutional Neural
Network  blocks, followed by a 2-layer Transformer encoder and a 1-layer bidirectional Long Short-Term Memory network. 
For the classification component, $3$ linear layers are used to predict frame-level counterfeit detection labels, manipulation type, and deepfake source, respectively. 
The predicted frame-level counterfeit probabilities are subsequently processed with median filtering.

The model was trained for $10$ epochs using the AdamW optimizer with a learning rate of $10^{-3}$. 
The encoder is frozen.
The frame-level detection task employed Binary Cross-Entropy loss, while the manipulation type and source classification tasks used Cross-Entropy loss. 
The total loss was formulated as a weighted sum of these losses with coefficients of $0.5$, $0.01$, and $0.01$.

\subsection{Evaluation Metric}
Following ADD2023~\cite{yi2023add}, we employed F1$_{segment}$ to evaluate frame-level predictions.
If the model predicts all frames within a clip as genuine, the clip-wise label is predicted as authentic; conversely, it is labeled as forged.
The clip-wise identification accuracy, manipulation type accuracy, and deepfake source accuracy are denoted as Acc$_{identify}$, Acc$_{manipulation}$, and Acc$_{source}$, respectively.

\section{Result and Analysis}
The results presented in Table~\ref{tab:result} demonstrate that, overall, current models exhibit strong localization capabilities but perform poorly in explainability, particularly when dealing with unseen sources.

\paragraph*{Localization}
The model demonstrates strong localization capabilities, as evidenced by its high performance on metric Acc$_{identify}$ and F1$_{segment}$. 
This indicates its effectiveness in accurately identifying and temporally localizing forged segments within the audio.
A similar conclusion can be drawn from the visualization in Figure~\ref{fig:feature_visualization}: the features extracted by the model exhibit considerable divergence in their spatial distributions between authentic and forged audio, resulting in strong binary classification performance. 
Meanwhile, for categories with authentic source audio (such as genuine, addition, and splice), their feature representations are closely clustered, indicating that the model has, to some extent, learned a robust concept of ``authenticity”.

\paragraph*{Explainability}
Despite proficient localization, the model exhibits notable limitations in explainability. 
Metrics such as Acc$_{source}$ and Acc$_{manipulation}$ remain relatively low across certain tasks, indicating an inability to meaningfully distinguish between manipulation types or attribute forgeries to specific sources. 
For example, the Editing-type manipulations are frequently misclassified as Generation-type forgeries. 
This shortcoming underscores the challenge of differentiating highly similar manipulation techniques and sources within the proposed dataset. 
Figure~\ref{fig:feature_visualization} reveals that forged audio categories with similar task objectives (e.g., Editing versus Generation) or model architectures (e.g., Affusion versus Audioeditor) exhibit entangled representations in the latent space. 
This suggests that the model fails to learn semantically meaningful distinctions between them.


\paragraph*{Generalization}
The performance  exhibits significant degradation on the majority of OOD tasks.
As illustrated in Figure~\ref{fig:feature_visualization}(b), the feature representations of X2Audio—a typical example—appear highly scattered and nearly random, leading to significantly lower detection accuracy compared to in-domain evaluations. 
These observations indicate that the current model fails to capture the underlying distribution of forged audio; rather, it depends on recognizing training-set-specific artifacts, resulting in poor generalization capability.
Therefore, we have identified several key challenges in current explainable deepfake sound detection:
(1) Capturing fine-grained distributional discrepancies arising from similar manipulation types or homogenized model architectures;
(2) Enabling models to characterize the true data distribution of authentic audio rather than memorizing artificial forgery patterns, thereby enhancing generalization capability.


\section{Conclusion}
This paper addresses a gap in deepfake sound detection by introducing a new benchmark specifically designed for explainability and generalization in synthetic audio identification.
The dataset aims to encourage further research within the community, raise awareness of ethical and security concerns in AIGC (AI-Generated Content), and provide justifiable guarantees for the outcomes of detection models.
Through a structured data pipeline, we constructed $6$ distinct manipulation types, along with $12$ manipulation sources, to provide diverse sources of forged audio segments, thereby facilitating model training focused on source attribution. 
Results show that current models perform well on real / fake classification and fake area localization. 
However, they struggle with traceability tasks. 
This suggests they do not truly understand fake audio content.
We contend that the primary challenge in DSD lies in (1) discerning fine-grained discrepancies among similar manipulation types and sources, while also (2) ensuring that models learn the underlying distribution of genuine audio rather than overfitting on artifactual patterns from forged training samples to enhance generalization.
We hope this work establishes a solid foundation and offers valuable insights to the field of DSD.

\bibliographystyle{IEEEbib}
\bibliography{strings,refs}

\end{document}